\documentclass[aps,prl,twocolumn,floatfix,citeautoscript,hyperref=pdftex]{revtex4-1}
\usepackage{graphicx}
\usepackage{dcolumn}
\usepackage{bm}
\usepackage{color}
\usepackage{amsmath}
\usepackage{amssymb}
\newcommand {\beq} {\begin{equation}}
\newcommand {\eeq} {\end{equation}}
\newcommand {\bqa} {\begin{eqnarray}}
\newcommand {\eqa} {\end{eqnarray}}

\usepackage[colorlinks=true]{hyperref}

\begin{document}

\title{Quasiparticle interference as a direct experimental probe of bulk odd-frequency superconducting pairing}

\author{Debmalya Chakraborty}
\affiliation{Department of Physics and Astronomy, Uppsala University, Box 516, S-751 20 Uppsala, Sweden}

\author{Annica M. Black-Schaffer}
\affiliation{Department of Physics and Astronomy, Uppsala University, Box 516, S-751 20 Uppsala, Sweden}

\begin{abstract}

We show that quasiparticle interference (QPI) due to omnipresent weak impurities and probed by Fourier transform scanning tunneling microscopy and spectroscopy acts as a direct experimental probe of bulk odd-frequency superconducting pairing. Taking the example of a conventional $s$-wave superconductor under applied magnetic field, we show that the nature of the QPI peaks can only be characterized by including the odd-frequency pairing correlations generated in this system. In particular, we identify that the defining feature of odd-frequency pairing gives rise to a bias asymmetry in the QPI, present generically in materials with odd-frequency pairing irrespective of its origin.

\end{abstract}

\maketitle

The key to finding new superconductors lies in understanding the nature of the Cooper pairs as they are the fundamental building blocks of the superconducting state. In particular, the symmetry of the Cooper pair wave function is crucial in determining the superconductor's stability, both to intrinsic constituents such as disorder and to external perturbations like magnetic field. The earliest known metallic superconductors have been successfully described by the Bardeen, Cooper, and Schrieffer (BCS) theory \cite{Bardeen57}, which is built on the Cooper pair wave function having spin-singlet $s$-wave symmetry and also an even-frequency dependence or, equivalently, being even under the exchange of the relative time coordinate of the paired electrons. Even after the discovery of high-temperature superconductors, the overall paradigm of BCS theory of superconductivity have remained fairly successful, albeit using other spin and spatial symmetries \cite{Tsuei00,Chubukov15}. However, there exist several properties in superconductor-ferromagnetic heterostructures \cite{Bergeret01,Bergeret01prb,Buzdin05,Eschrig08,DiBernardo15,Bernardo15,Jacobsen16,Ouassou17,Perrin20}, which can only be explained by the presence of Cooper pairs with a wave function that is odd in frequency \cite{Berezinskii74, Kirkpatrick91, Balatsky92, Schrieffer94, Bergeret05,Yokoyama11,Alidoust14,Tanaka12,Linder19}.

Experimental evidence of odd-frequency pairing in superconductor-ferromagnetic heterostructures \cite{Bergeret01,Bergeret01prb,Buzdin05,Eschrig08,DiBernardo15,Bernardo15,Krieger20} have motivated other theoretical works, proposing the existence of odd-frequency pairs also in bulk systems without the need of heterostructures \cite{Triola18,Chakraborty21,Dutta21,Cayao21,Black-Schaffer13,Komendova15,Asano15,Komendova17,Tsvelik19,Triola20,Schmidt20,deCarvalho21,Yoshida21}. Additionally, bulk odd-frequency pairing has also been discussed theoretically \cite{Coleman93,Fuseya03,Otsuki15,Wu22} and experimentally \cite{Kawasaki20} in the context of heavy fermions for pairing near quantum critical points. However, a major challenge has been the experimental verification of such theoretical proposals of bulk odd-frequency superconducting pairs. While experiments have been successful in identifying odd-frequency pairing in heterostructures \cite{Bergeret01, Buzdin05, Eschrig08, DiBernardo15, Bernardo15, Krieger20}, there exist still no proposal of an easily accessible experimental probe which clearly identifies odd-frequency Cooper pairs in bulk systems. The challenge mainly arises due to the fact that  bulk odd-frequency pairs are generally accompanied by even-frequency pairs and it is difficult to experimentally  disentangle their different characteristics. A few proposals still exist, such as using the Kerr effect \cite{Komendova17} or a paramagnetic Meissner effect \cite{Bergeret01prb,Yokoyama11,Alidoust14,Bernardo15,Krieger20}. However, the Kerr effect additionally requires time-reversal symmetry breaking \cite{Xia06,Kapitulnik09} and the Meissner effect has been shown to be unreliable in multiband superconductors since odd-frequency pairing can also generate a diamagnetic Meissner signal \cite{Schmidt20,Parhizgar21}. Additionally, neither of these tools directly detect the oddness in frequency, but instead rely on indirect effects.
Another theoretical proposal has tried to provide a direct experimental detection scheme of bulk odd-frequency Cooper pairs using time- and angle-resolved photoelectron fluctuation spectroscopy \cite{Kornich21}, but involves technology clearly beyond the scope of even the most advanced existing facilities.

In this Letter we show that the already existing experimental technique of scanning tunneling microscopy/spectroscopy (STM/STS) can directly detect odd-frequency superconducting pairing.  In particular, we use that weak non-magnetic impurities in superconductors are ever-present and create charge density inhomogeneities, which result in quasiparticle scattering. The resulting interference patterns can be probed experimentally by Fourier transformed STM/STS in a technique commonly referred to as quasiparticle interference (QPI) \cite{Hoffman02a, Wang03, McElroy03}. QPI has long been an important experimental probe to determine various signatures of superconducting pairing \cite{Hanaguri09,Liu19,Cheung20,Boker20}, especially the pairing symmetry in high-temperature superconductors such as iron-based superconductors \cite{Hirschfeld15,Altenfeld18,Du18,Sharma21}. Here, by considering a prototype system of a conventional spin-singlet $s$-wave superconductor under applied magnetic field, we show that odd-frequency pairing, known to be present in such systems \cite{Eschrig07,Eschrig15,Spintronicsbook}, produces two direct signatures in the peak structure of the QPI. First, we show that the peak positions of LDOS-change can only be identified accurately if odd-frequency pair correlations are incorporated. Secondly and most remarkably, the LDOS-change at positive and negative applied bias voltage results in different peak positions, and their separation is directly related to the presence of odd-frequency pairs. This bias asymmetry arises due to the defining feature of odd-frequency pairing, i.e., the superconducting pair correlations being odd in frequency or, equivalently, here in applied bias voltage.

{\it{Model.}}-- To model a simple bulk superconductor with odd-frequency superconducting pair correlations we use a conventional spin-singlet $s$-wave superconductor under applied magnetic field described by the mean-field Hamiltonian: 
\begin{eqnarray}
H=\sum_{k,\sigma} \left(\xi_{k}+\sigma B \right) c_{k \sigma}^{\dagger} c_{k \sigma} 
+ \sum_{k} \Delta_{0} c_{-k \downarrow} c_{k \uparrow} + \textrm{H.c.}
\label{eq:Hamil}
\end{eqnarray}
where $c_{k \sigma}^{\dagger}$ ($c_{k \sigma}$) is the creation (annihilation) operator of an electron with spin $\sigma$ and momentum $k$, $\xi_{k}$ is the electron band dispersion, $B$ is the magnetic field with the magnetic moment of the electron $\mu_0$ taken to be unity, $\Delta_0$ is the spin-singlet isotropic $s$-wave superconducting order parameter, and we have ignored an overall constant. For simplicity, we use the band dispersion of a square lattice: $\xi_{k}=-2t(\cos(k_x)+\cos(k_y))-\mu$, where $t=1$ is the energy unit. We only consider the Zeeman effect of the applied magnetic field in this work. An applied magnetic field can also affect the orbital motion of electrons and create vortices in a superconductor. However, in two-dimensional superconductors with the magnetic field applied in-plane, the only relevant effect is the Zeeman effect. Furthermore, $\Delta_0$ is obtained by the self-consistency relation, $\Delta_0=-\sum_{k^{\prime}}U \langle c_{k^{\prime} \uparrow}^{\dagger} c_{-k^{\prime} \downarrow}^{\dagger} \rangle$ with $U$ being the effective attraction driving the superconducting order. 

The Hamiltonian in Eq.~\eqref{eq:Hamil} can be written in a matrix form using the Nambu basis $\Psi^{\dagger}=\left(c_{k \uparrow}^{\dagger},c_{-k \downarrow}\right)$ as,
$H=\sum_{k} \Psi^{\dagger} \hat{H} \Psi$
with
\begin{equation}
\hat{H}=\left(\begin{array}{cc} \xi_{k\uparrow} & \Delta_{0} \\
\Delta_{0} & -\xi_{-k\downarrow} \\
\end{array}\right),
\label{eq:Hamilmat}
\end{equation} 
where now $\xi_{k \sigma}=\xi_{k}+\sigma B$ and $\Delta_{0}$ is, without loss of generality, taken to be real. We diagonalize the Hamiltonian $\hat{H}$ and solve the self-consistent equation of $\Delta_0$ iteratively. We also tune $\mu$ such that the average density of electrons $\rho=\sum_{k,\sigma}\langle c^{\dagger}_{k\sigma}c_{k\sigma}\rangle$ is kept fixed to the generic value $0.7$. Our findings do not qualitatively depend on the choice of the average density. We use a large system size, $N=1000\times 1000$ and $U=2.5$ for obtaining a significant gap $\Delta_0$ to make the analysis clear. We have verified that all qualitative features remain for experimentally realistic gap sizes. For the given set of parameters, we find the BCS superconducting state to be stable for $B<0.35$. 

{\it{Odd-frequency pair correlations.}}-- Different correlations of the Hamiltonian in Eq.~\eqref{eq:Hamilmat} can be obtained by calculating the corresponding Green's function $G$, given by $G^{-1}(i\omega)=i\omega-\hat{H}$, where $\omega$ are fermionic Matsubara frequencies. Using the Hamiltonian in Eq.~\eqref{eq:Hamilmat}, we thus obtain the Green's function by inverting the $2\times2$ matrix $G^{-1}(i\omega)$. In particular, the superconducting pair correlations is given by the anomalous, off-diagonal, part of the Green's function,
\begin{equation}
G_{12}(i\omega)=F_k(i\omega)=F^{e}_{k}(i\omega)+F^{o}_{k}(i\omega),\label{eq:anomolous}
\end{equation}
where we analytically extract
\begin{eqnarray}
&&F^{e}_{k}(i\omega)=\frac{-\Delta_0\left( \xi_{k\uparrow}\xi_{-k\downarrow}+\Delta_0^2+\omega^2 \right)}{D},\label{eq:feven}\\
&&F^{o}_{k}(i\omega)=\frac{i\omega\Delta_0\left( \xi_{k\uparrow}-\xi_{-k\downarrow} \right)}{D},\label{eq:fodd} \\
&&D=\left( \xi_{k\uparrow}\xi_{-k\downarrow}+\Delta_0^2+\omega^2 \right)^2+\omega^2\left( \xi_{k\uparrow}-\xi_{-k\downarrow} \right)^2. \label{eq:D}
\end{eqnarray}
Here we have divided the superconducting pair correlations into even-frequency, $F^{e}_{k}(i\omega)$, and odd-frequency, $F^{o}_{k}(i\omega)$, contributions, as clearly set by the frequency dependence in the numerators since the common denominator $D$ is an even function of frequency.  Moreover, we see directly that odd-frequency pairs exist as soon as $B$ is finite, as expected with a triplet spin symmetry \cite{Eschrig07,Eschrig15,Spintronicsbook}.

{\it{QPI theory.}}-- QPI probes the change in the LDOS due to omnipresent weak non-magnetic impurities. The LDOS in the presence of such impurities can be decomposed as $\rho(r,\omega)=\rho_0(\omega)+\delta\rho(r,\omega)$, where $\rho_0$ is the DOS of a homogeneous superconductor and $\delta\rho$ is the change due to impurities. The corresponding Fourier transformed quantity $\delta\rho(q,\omega)$ is written in terms of the Green's function as \cite{Hirschfeld15,Altenfeld18,Liu19,Cheung20,Boker20,Sharma21},
\begin{equation}
\delta\rho(q,\omega)=-\frac{1}{\pi}Im\left[\sum_{k}G(k,\omega)TG(k+q,\omega)\right]_{11},
\label{eq:rhoqdef}
\end{equation}
where $G(k,\omega)$ is obtained by analytically continuing $i\omega\rightarrow\omega+i\eta$ in the unperturbed $G(k,i\omega)$ and $T$ is the T-matrix \cite{MahanBook} corresponding to the impurity. Assuming weak non-magnetic impurities, $T=V_{\text {imp}}\tau_3$ \cite{Hirschfeld15}, where $\tau_3$ is the third Pauli matrix in the Nambu basis and we set $V_{\text {imp}}=1$ to mimic weak impurities.
This impurity treatment is sufficient as the superconducting properties or the ground state are not changed in an $s$-wave superconductor for weak non-magnetic impurities due to the Anderson's theorem \cite{Anderson59}. Moreover, weak non-magnetic impurities do not create additional local odd-frequency correlations \cite{Triola19}, thus helping in unambiguous detection of bulk odd-frequency correlations. Strong or magnetic impurities may create additional bound states. However, experimentally such bound states are often easily isolated and subtracted from the signal obtained for QPI \cite{Du18}. The quantity $\delta\rho(q,\omega)$ is also experimentally accessible by Fourier transforming the LDOS obtained using STM/STS at different bias voltages $V=\omega/e$, where $e$ is the electron charge \cite{Hirschfeld15}. 

For weak non-magnetic impurities, Eq.~\eqref{eq:rhoqdef} becomes
\begin{subequations}
\begin{eqnarray}
\delta\rho(q,\omega)&&=-\frac{1}{\pi}Im\Big[\sum_{k}G_{11}(k,\omega)G_{11}(k+q,\omega) \nonumber \\
&&-G_{12}(k,\omega)G_{21}(k+q,\omega)\Big], \nonumber \\
&&=-\frac{1}{\pi}Im\Big[\sum_{k}G_{11}(k,\omega)G_{11}(k+q,\omega) \label{eq:rhoqdeffefoa} \\
&&-F_k^{e}(\omega)F_{k+q}^{e}(\omega)-F_k^{o}(\omega)F_{k+q}^{o}(\omega) \label{eq:rhoqdeffefob} \\
&&-F_k^{e}(\omega)F_{k+q}^{o}(\omega)-F_k^{o}(\omega)F_{k+q}^{e}(\omega)\Big] \label{eq:rhoqdeffefoc},
\end{eqnarray}
\label{eq:rhoqdeffefo}
\end{subequations}
where $F_k^{e}(\omega)$ and $F_k^{o}(\omega)$ are analytically continued versions of $F_k^{e}(i\omega)$ and $F_k^{o}(i\omega)$, respectively. Looking at the expressions in Eq.~\eqref{eq:rhoqdeffefo}, we can already make two key observations. First, $\delta\rho(q,\omega)$ directly access the pair correlations through the terms proportional to $F^{e}_{k}$ and $F^{o}_{k}$. This is in sharp contrast to the homogeneous DOS $\rho_{0}(\omega)=-1/\pi \sum_{k}Im[G_{11}(k,\omega)]$, which only captures the superconducting energy gap $\Delta_0$ through $G_{11}$ and can thus not probe the pair correlations $F^{e/o}_{k}$. 
Secondly, the expression Eq.~\eqref{eq:rhoqdeffefo} has no integration over $\omega$. As a result, the contributions coming from the product of the even- and odd-frequency correlations are generically non-zero. These contributions are odd in $\omega$ by definition, in contrast to the contributions coming from so-called even-even $F_k^{e}(\omega)F_{k+q}^{e}(\omega)$ or odd-odd $F_k^{o}(\omega)F_{k+q}^{o}(\omega)$ correlations. Hence, we already here find that the presence of odd-frequency correlations generically, irrespective of its origin, influences $\delta\rho(q,\omega)$.

{\it{QPI results.}}-- In order to identify the role of odd-frequency correlations in the LDOS-change $\delta\rho(q,\omega)$, we compare $\delta\rho(q,\omega)$ with $\delta\rho_e(q,\omega)$, where $\delta\rho_e(q,\omega)$ is calculated altogether ignoring odd-frequency correlations and keeping only even-frequency correlations, i.e.~$F_k^{o}(\omega)$ is set to zero in the expression in Eq.~\eqref{eq:rhoqdeffefo} to obtain $\delta\rho_e(q,\omega)$. We here present $\omega$ is in units of $t$ and  the LDOS-change in units of inverse $t$.

\begin{figure}[t]
\includegraphics[width=1.0\linewidth]{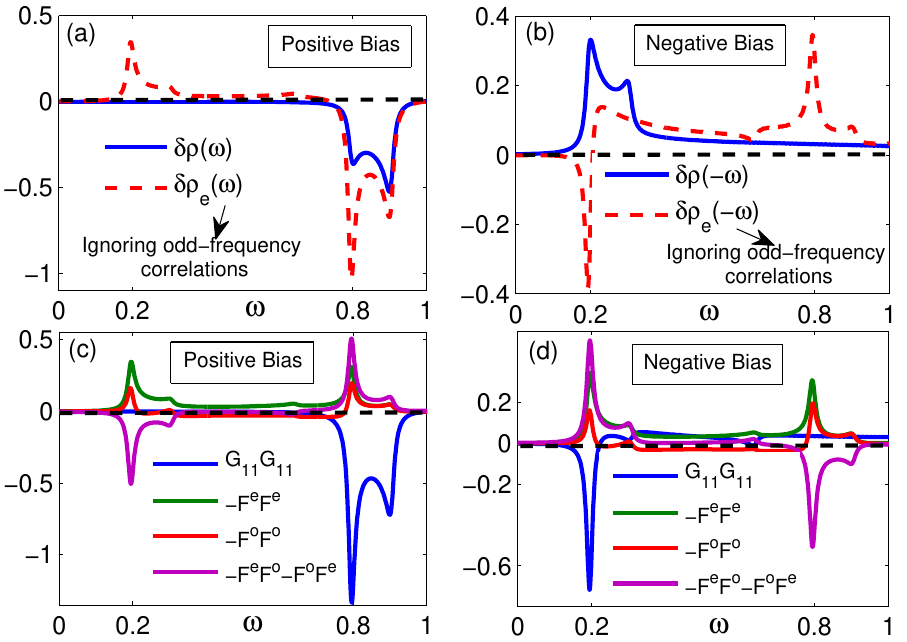} \caption{LDOS-change as a function of $\omega$ for $B=0.3$ and $q=(1.37,0)$, showing different peak structures. $\delta\rho$ and $\delta\rho_e$ for positive bias (a) and negative bias (b). Individual contributions of each term in Eq.~\eqref{eq:rhoqdeffefo} to $\delta\rho$ for positive bias (c) and negative bias (d). Black dashed line marks zero. Results are obtained for an artificial broadening $\eta=0.01$.}
\label{fig:diffpeaks} 
\end{figure}

In Fig.~\ref{fig:diffpeaks}(a,b) we show the LDOS-changes $\delta\rho(q,\omega)$ and $\delta\rho_e(q,\omega)$ considering the system in Eq.~\eqref{eq:Hamil} as a function of $\omega$ for a fixed $q=(1.37,0)$ and $B=0.3$. The choice of $B$ is set to give significant odd-frequency correlations, as also illustrated later in Fig.~\ref{fig:summary}. We show in the Supplementary Material (SM) \footnote{See Supplementary Material where we show that the features presented in the main text are generic and not dependent on specific parameters or other symmetries of the Cooper pairs, also including Refs.~\onlinecite{Waardh17,SudboBook}.} that the qualitative features obtained in Fig.~\ref{fig:diffpeaks} are indeed independent of the choice of $q$ or $B$. Focusing on the positive bias ($\omega>0$) in (a), we find that $\delta\rho(\omega)$ has a negative double-peak structure around $\omega=0.8=\Delta_0+B$. When ignoring odd-frequency correlations, $\delta\rho_e(\omega)$ shows also an additional peak at $\omega=0.2=\Delta_0-B$. The lack of this spurious peak in the full $\delta\rho(q,\omega)$ is the first evidence of the presence of odd-frequency correlations. If we look at the negative bias $\delta\rho_e(-\omega)$ in (b), we find a similar spurious peak at $\omega=0.8$. The negative bias results show further qualitative differences between $\delta\rho$ and $\delta\rho_e$. Around $\omega=0.2$, $\delta\rho(-\omega)$ has positive double-peak. In contrast, $\delta\rho_e(-\omega)$ has a sign-change occurring near $\omega=0.2$, such that $\delta\rho_e(-\omega)<0$ for $\omega<0.2$ and $\delta\rho_e(-\omega)>0$ for $\omega>0.2$. These results illustrate clearly that the LDOS-change as a function of $\omega$ attains a different peak structure due to the presence of odd-frequency correlations. 

The appearance of spurious peaks in $\delta\rho_e$ in contrast to the observable $\delta\rho$ for both positive and negative bias in Figs.~\ref{fig:diffpeaks}(a,b) can be understood by looking at the individual components of $\delta\rho$ in Eq.~\eqref{eq:rhoqdeffefo}. In Figs.~\ref{fig:diffpeaks}(c,d) we show the contributions coming from the normal, diagonal, part of the Green's function $G_{\rm{11}}G_{\rm{11}}=-Im(\sum_{k}G_{11}(k,\omega)G_{11}(k+q,\omega))/\pi$, i.e.~the term in Eq.~\eqref{eq:rhoqdeffefoa}, and the anomalous or superconducting, off-diagonal, part of the Green's function $F^{\rm{e/o}}F^{\rm{e/o}}=-Im(\sum_{k}F_k^{e/o}(\omega)F_{k+q}^{e/o}(\omega))/\pi$, i.e.~the terms in Eqs.~\eqref{eq:rhoqdeffefob} and \eqref{eq:rhoqdeffefoc}. In particular, we note that the contribution from $F^{\rm{e}}F^{\rm{o}}+F^{\rm{o}}F^{\rm{e}}$ exactly cancels $F^{\rm{e}}F^{\rm{e}}+F^{\rm{o}}F^{\rm{o}}$ at $\omega=0.2$ for positive bias in (c) and at $\omega=0.8$ for negative bias in (d). These exact cancellations can be easily seen by comparing the total $\delta\rho$ in Figs.~\ref{fig:diffpeaks}(a,b) with the normal part $G_{\rm{11}}G_{\rm{11}}$ in Figs.~\ref{fig:diffpeaks}(c,d). Moreover, $G_{\rm{11}}G_{\rm{11}}$ do not feature any peaks at these $\omega$ values. As a result, it is only in the presence of odd-frequency correlations, i.e.~when $F^{o}\ne 0$, that there are no spurious peaks at $\omega=0.2$ for positive bias and $\omega=0.8$ for negative bias in $\delta\rho$.

\begin{figure}[t]
\includegraphics[width=1.0\linewidth]{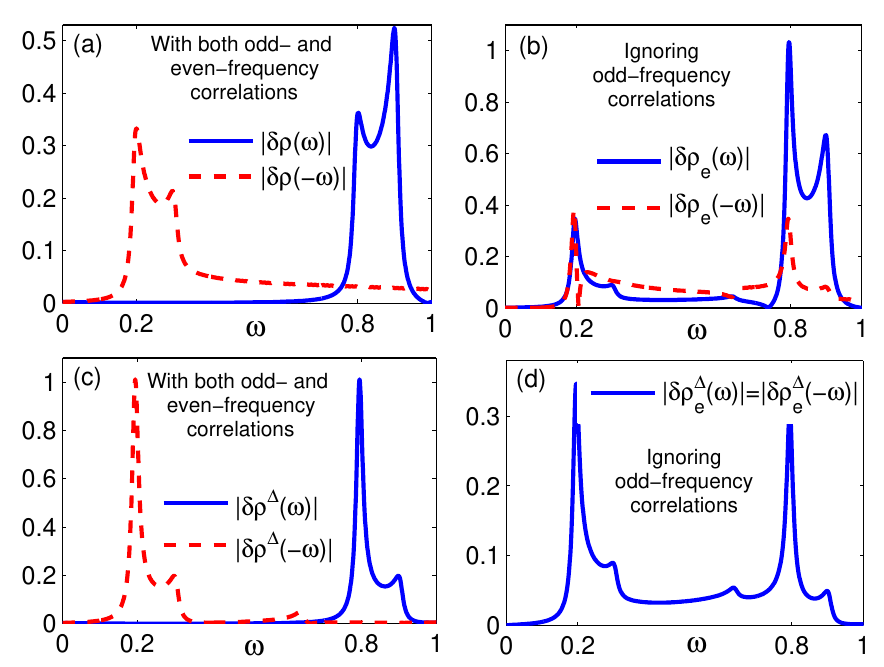} \caption{Absolute value LDOS-change as a function of $\omega$ for $B=0.3$ and $q=(1.37,0)$, showing bias asymmetry. Positive and negative bias values of $|\delta\rho|$ (a) and $|\delta\rho_{e}|$ (b), as well as the contributions only from superconductivity $|\delta\rho^{\Delta}(\pm \omega)|$ (c) and $|\delta\rho^{\Delta}_{e}(\pm \omega)|$ (d). Artificial broadening is same as in Fig.~\ref{fig:diffpeaks}.}
\label{fig:posneg} 
\end{figure}

Further looking for signatures of odd-frequency correlations, we reemphasize that the defining feature of odd-frequency correlations is that they are odd in frequency or, equivalently, energy bias. Therefore, we particularly want to explore the possibility of a bias asymmetry in the LDOS-change. For this purpose, we plot the absolute values $|\delta\rho(\omega)|$ and $|\delta\rho(-\omega)|$ in Fig.~\ref{fig:posneg}(a), and for comparison $|\delta\rho_e(\omega)|$ and $|\delta\rho_e(-\omega)|$ in Fig.~\ref{fig:posneg}(b), again using a fixed $q=(1.37,0)$ with $B=0.3$. In Fig.~\ref{fig:posneg}(a), $|\delta\rho(\omega)|$ shows the double-peak centered around $\omega=0.8$, whereas the double-peak of $|\delta\rho(-\omega)|$ is centered around $\omega=0.2$. This illustrates directly a clear asymmetry in the peak positions of $|\delta\rho(\omega)|$ and $|\delta\rho(-\omega)|$ when odd-frequency correlations are appropriately included. In comparison, the LDOS-change if ignoring odd-frequency correlations are shown in (b). It is evident from (b) that both $|\delta\rho_e(\omega)|$ and $|\delta\rho_e(-\omega)|$ have peaks around both $\omega=0.2$ and $\omega=0.8$. Thus, there are no asymmetry in the positions of the peaks of $|\delta\rho_e(\omega)|$ and $|\delta\rho_e(-\omega)|$. As a consequence, the bias asymmetry in the peak positions of $|\delta\rho(\omega)|$ and $|\delta\rho(-\omega)|$ becomes a clear and robust experimental signature of the presence of odd-frequency correlations. However, we still note that the heights of the peaks for positive and negative bias are asymmetric in both (a) and (b). To analyze this peak height asymmetry, in (c) and (d) we plot the LDOS-change after subtracting the contribution coming from the normal part of the Green's function, resulting in the isolation of the contribution from superconducting correlations, $\delta\rho^{\Delta}=\delta\rho-G_{\rm{11}}G_{\rm{11}}$. If the odd-frequency correlations are ignored, as is shown in (d), then the positive and negative bias results are exactly identical, i.e.~$|\delta\rho^{\Delta}_e(\omega)| = |\delta\rho^{\Delta}_e(-\omega)|$. This is an expected results since the only contribution to $\delta\rho^{\Delta}$ comes from $F^eF^e$, which is an even function in frequency. Moreover, once the odd-frequency correlations are appropriately included, we find that $\delta\rho^{\Delta}(\omega)$ and $\delta\rho^{\Delta}(-\omega)$ peaks at $\omega=0.8$ and $\omega=0.2$, respectively, and the peak heights are in fact identical. Thus, the asymmetry in the peak heights in Figs.~\ref{fig:posneg}(a,b) is stemming from the contribution of the normal part of the Green's function. As a consequence, it is the bias asymmetry in the peak positions that is the decisive tool to prove odd-frequency pairing, while the heights are varying with the normal, non-superconducting properties. In the SM we show that these are robust features, not dependent on the choice of $q$ or $B$. Another striking feature is that the bias asymmetry only appears in the LDOS-change, or QPI, and not in any spatially averaged tunneling measurements, as the average DOS does not capture the superconducting correlations, as mentioned earlier.
Hence, averaged tunneling measurements see two symmetric peaks at $\omega=\Delta_0\pm B$ for both positive and negative bias \cite{Meservey70,Spintronicsbook}.

\begin{figure}[t]
\includegraphics[width=1.0\linewidth]{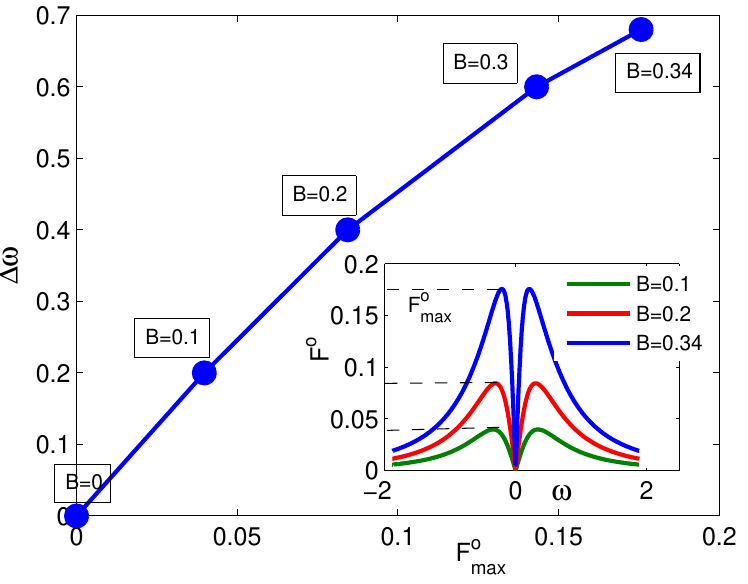} \caption{Correlation between the energy separation $\Delta \omega$ between the peaks of $|\delta\rho^{\Delta}(\omega)|$ and $|\delta\rho^{\Delta}(-\omega)|$ and maximum value of the momentum averaged odd-frequency pair correlations $F^{o}_{\rm{max}}$ for multiple different $B$ (indicated at each point). Inset: $F^{o}$ for different values of magnetic field $B$, with $F^{o}_{\rm{max}}$ values indicated by dashed black lines.}
\label{fig:summary} 
\end{figure}

In order to further justify the bias asymmetry originating from odd-frequency pair correlations, we show the dependence of the odd-frequency correlations on the applied magnetic field $B$ in the inset of Fig.~\ref{fig:summary}. Here we plot the momentum integrated values $F^{o}(\omega)=\sum_{k}|Im{F^{o}_{k}}|$ for three different values of B. $F^{o}$ is clearly zero at $\omega=0$ and peaks to a maximum value $F^{o}_{\rm{max}}$ at a finite $\omega$. With increasing $B$, $F^{o}_{\rm{max}}$ also increases. This increase can easily be understood by looking at the expression for $F^{o}_{k}(i\omega)$ in Eq.~\eqref{eq:fodd}. The numerator of Eq.~\eqref{eq:fodd} depends linearly on $\xi_{k\uparrow}-\xi_{k\downarrow}=2B$ and it also primarily decides the maximum value $F^{o}_{\rm{max}}$. To relate these odd-frequency pair correlations with the bias asymmetry we uncovered in Fig.~\ref{fig:posneg}, we first note that the distance $\Delta\omega$ between the peaks of $\delta\rho^{\Delta}(\omega)$ and $\delta\rho^{\Delta}(-\omega)$ is always equal to $2B$, since $\delta\rho^{\Delta}(\omega)$ peaks at $\omega=\Delta_0+B$ and $\delta\rho^{\Delta}(-\omega)$ peaks at $\omega=\Delta_0-B$. To numerically substantiate this result, we show in the main panel of Fig.~\ref{fig:summary} the relation between $\Delta\omega$ and $F^{o}_{\rm{max}}$ obtained at multiple different $B$. Importantly, as seen in Fig.~\ref{fig:summary}, $\Delta\omega$ is directly correlated to $F^{o}_{\rm{max}}$. This correlation establishes the direct connection of the odd-frequency correlations and the bias asymmetry in LDOS-change. The correlation plot in Fig.~\ref{fig:summary} is not quite linear because the $B$-dependence of $F^{o}_{\rm{max}}$ is not exactly linear due to also a $B$-dependence in the denominator in Eq.~\eqref{eq:fodd}.

{\it{Concluding remarks.}}-- We showed that odd-frequency superconducting pair correlations can be directly probed by the quasiparticle interference (QPI) technique. Studying a conventional $s$-wave superconductor under applied magnetic field, we identified the relation between the odd-frequency superconducting pair correlations and different peaks in the change of the Fourier transformed local density of states (LDOS) directly measurable by STM/STS. Remarkably, we find a direct relationship between odd-frequency correlations and a bias asymmetry in the LDOS-change. 
This method is applicable irrespective of the other symmetries of the Cooper pairs, due to it explicitly probing the oddness in frequency. We further validate this in the SM by showing the same LDOS-change in a superconductor with spin-singlet $p$-wave odd-frequency pair correlations \cite{Chakraborty22}.
Similar QPI analysis can also be performed in other materials with bulk odd-frequency correlations, for example multiband superconductors \cite{Black-Schaffer13,Komendova15,Asano15,Komendova17,Triola20,Schmidt20}, Ising superconductors \cite{Tang21}, non-Hermitian superconductors \cite{Cayao21}, heavy fermion \cite{Kawasaki20}, spin-3/2 systems with Bogoliubov Fermi surfaces, \cite{Dutta21} and also in heterostructures to directly verify the existence of odd-frequency correlations. In fact, relevant QPI peaks, especially a bias asymmetry, have already been observed in heavy fermion \cite{Zhou13} and iron-based multiband \cite{Du18,Sharma21} superconductors, in the latter explained in terms of a sign-changing order parameter \cite{Du18,Sharma21}, but with no current consensus and with the possibility of odd-frequency not yet explored. Our work thus provides a pathway in characterizing the pairing symmetries in novel superconductors.

We gratefully acknowledge financial support from the Knut and Alice Wallenberg Foundation through the Wallenberg Academy Fellows program and the European Research Council (ERC) under the European Unions Horizon 2020 research and innovation programme (ERC-2017-StG-757553). The computations were enabled by resources provided by the Swedish National Infrastructure for Computing (SNIC) at the Uppsala Multidisciplinary Center for Advanced Computational Science (UPPMAX) partially funded by the Swedish Research Council through grant agreement No.~2018-05973.

 \bibliographystyle{apsrev4-1}
\bibliography{Cuprates}

\pagebreak
\widetext
\clearpage 
\normalsize
~\vspace{0.2cm} 
\setcounter{equation}{0}
\setcounter{figure}{0}
\setcounter{table}{0}
\setcounter{page}{1}
\makeatletter
\renewcommand{\theequation}{S\arabic{equation}}
\renewcommand{\thefigure}{S\arabic{figure}}
\renewcommand{\figurename}{FIG.}

\begin{center}
{\large\bf Supplementary Material for \\
 ``Quasiparticle interference as a direct experimental probe of bulk odd-frequency superconducting pairing"}
\end{center}

\vspace{0.2cm}

In this Supplementary Material (SM), we show that the features of the LDOS-change presented in the main text are generic and not dependent on specific parameters or other symmetries of the Cooper pairs. We do this by showing the LDOS-change for a different $q$ value from the value used in the main text and using more $B$ values, as well as mapping the LDOS-change in the full Brillouin zone for the model considered in the main text. We also show the LDOS-change for a very different model which generates spin-singlet odd-frequency pair correlations. 

\section{QPI for more parameters than shown in main text}

\begin{figure}[htb]
\includegraphics[width=0.7\linewidth]{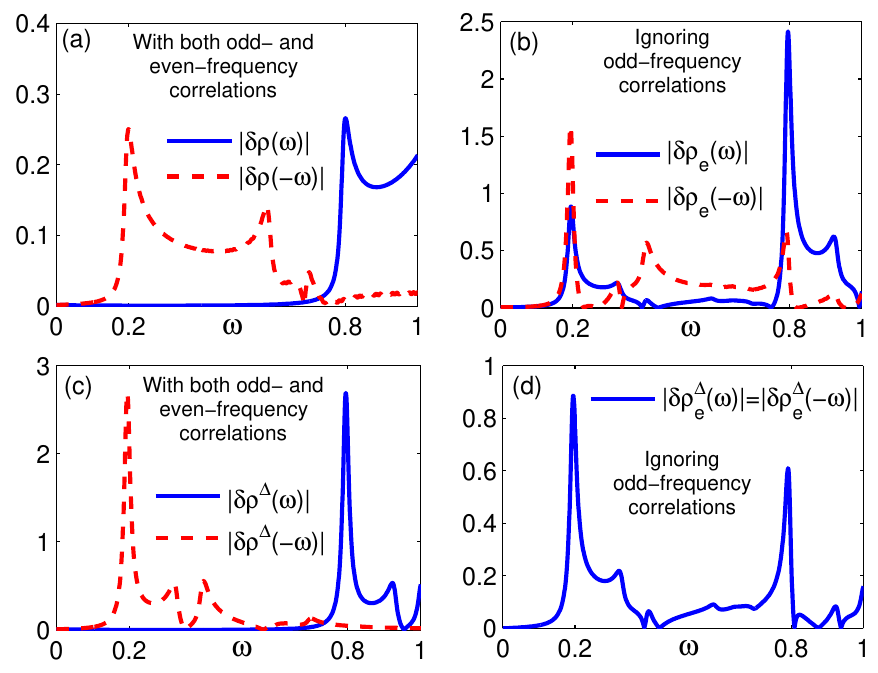} \caption{Absolute value LDOS-change as a function of $\omega$ for $B=0.3$ showing bias asymmetry, similar to Fig.~2 of the main text but here with $q=(0.5,0)$.
Positive and negative bias values of $|\delta\rho|$ (a) and $|\delta\rho_{e}|$ (b), as well as the contributions only from superconductivity $|\delta\rho^{\Delta}(\pm \omega)|$ (c) and $|\delta\rho^{\Delta}_{e}(\pm \omega)|$ (d). Artificial broadening is same as in Fig.~2 of the main text.}
\label{fig:posnegotherq} 
\end{figure}

In Fig.~2 of the main text we show the bias asymmetry in the LDOS-change for $q=(1.37,0)$. To show that our findings are not sensitive to this particular choice of $q$, we here in Fig.~\ref{fig:posnegotherq} plot the bias asymmetry for the same magnetic field $B$ but for a very different $q=(0.5,0)$. As seen, also here, $|\delta\rho(\omega)|$ and $|\delta\rho(-\omega)|$ peaks at very different $\omega$ in (a). In contrast, $|\delta\rho_e(\omega)|$ and $|\delta\rho_e(-\omega)|$ both have the same peak positions as shown in (b),  similar to the main text. Again, we show that the asymmetry in the peak heights are due to the normal part of the Green's function, as we see that the height asymmetry is not present in panels (c) and (d). The distance between the peaks in $\delta\rho^{\Delta}(\omega)$ and $\delta\rho^{\Delta}(-\omega)$ in (c) is again $2B$, similar to the Fig.~2 of the main text, even for a different $q=(0.5,0)$. Thus, by showing that the distance between the peak positions of $\delta\rho^{\Delta}(\omega)$ and $\delta\rho^{\Delta}(-\omega)$ is equal to $2B$ and also directly related to the odd-frequency correlations, for two different $q$ values, we argue that the direct connection between the bias asymmetry in the LDOS-change and odd-frequency correlations is a generic feature, independent of the choice of $q$.

We also note one qualitative difference of Fig.~2(a) of the main text and Fig.~\ref{fig:posnegotherq}(a). $|\delta\rho(-\omega)|$ in Fig.~2(a) has a double-peak structure around $\omega=0.2$. In contrast, $|\delta\rho(-\omega)|$ in Fig.~\ref{fig:posnegotherq}(a) has two peaks, one at $\omega=0.2$ and the other at $\omega=0.6$. However, this difference for $q=(1.37,0)$ and $q=(0.5,0)$ is not present if we only consider the contribution coming from the anomalous part of the Green's function, as shown in (c), and is thus generated by the normal-state properties.

\begin{figure*}[t]
\includegraphics[width=0.7\linewidth]{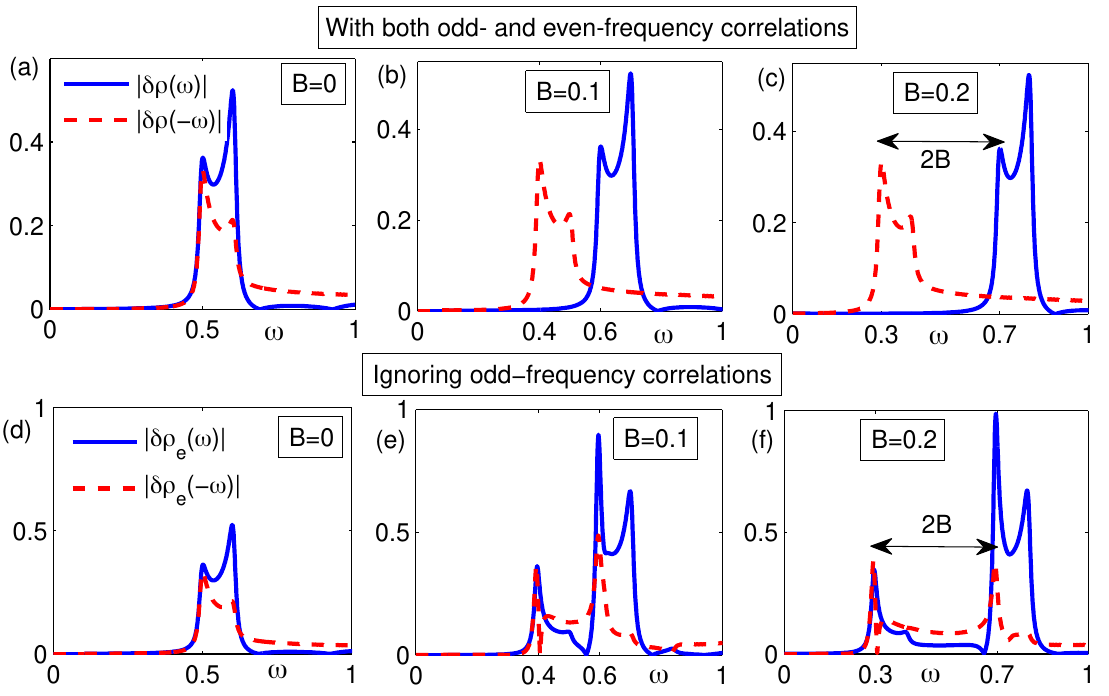} \caption{Bias asymmetry for different magnetic field $B$. (a-c) $|\delta\rho(\pm \omega)|$ after appropriately including odd-frequency correlations. (d-f) $|\delta\rho_e(\pm \omega)|$ where odd-frequency correlations are ignored.}
\label{fig:asfuncofB} 
\end{figure*}

In Fig.~2 of the main text we also only study a fixed value of magnetic field $B=0.3$ and show the connection of the bias asymmetry with the odd-frequency correlations. In order to further justify this connection, we plot the LDOS-change for several different magnetic fields $B$ in Fig.~\eqref{fig:asfuncofB} for a fixed $q=(1.37,0)$ (same as Fig.~2 in the main text). The peak positions of $|\delta\rho(\omega)|$ and $|\delta\rho(-\omega)|$ are found to be symmetric for $B=0$ as seen in (a). However, with increasing $B$ the asymmetry develops in the peak positions of the positive and negative bias LDOS-change, and the distance between the peaks is equal to $2B$ as seen in (b,c). Notably, ignoring odd-frequency correlations we do not find any bias asymmetry in the peak positions of $|\delta\rho_e(\omega)|$ and $|\delta\rho_e(-\omega)|$ for all $B$, as shown in (d-f).    
Similar to the peak position asymmetry, the odd-frequency correlations are also absent for $B=0$ and increases with increasing $B$. This remarkable similarity of the $B$-dependence directly relates bias asymmetry of LDOS-change to the odd-frequency correlations. 

\begin{figure}[htb]
\includegraphics[width=0.7\linewidth]{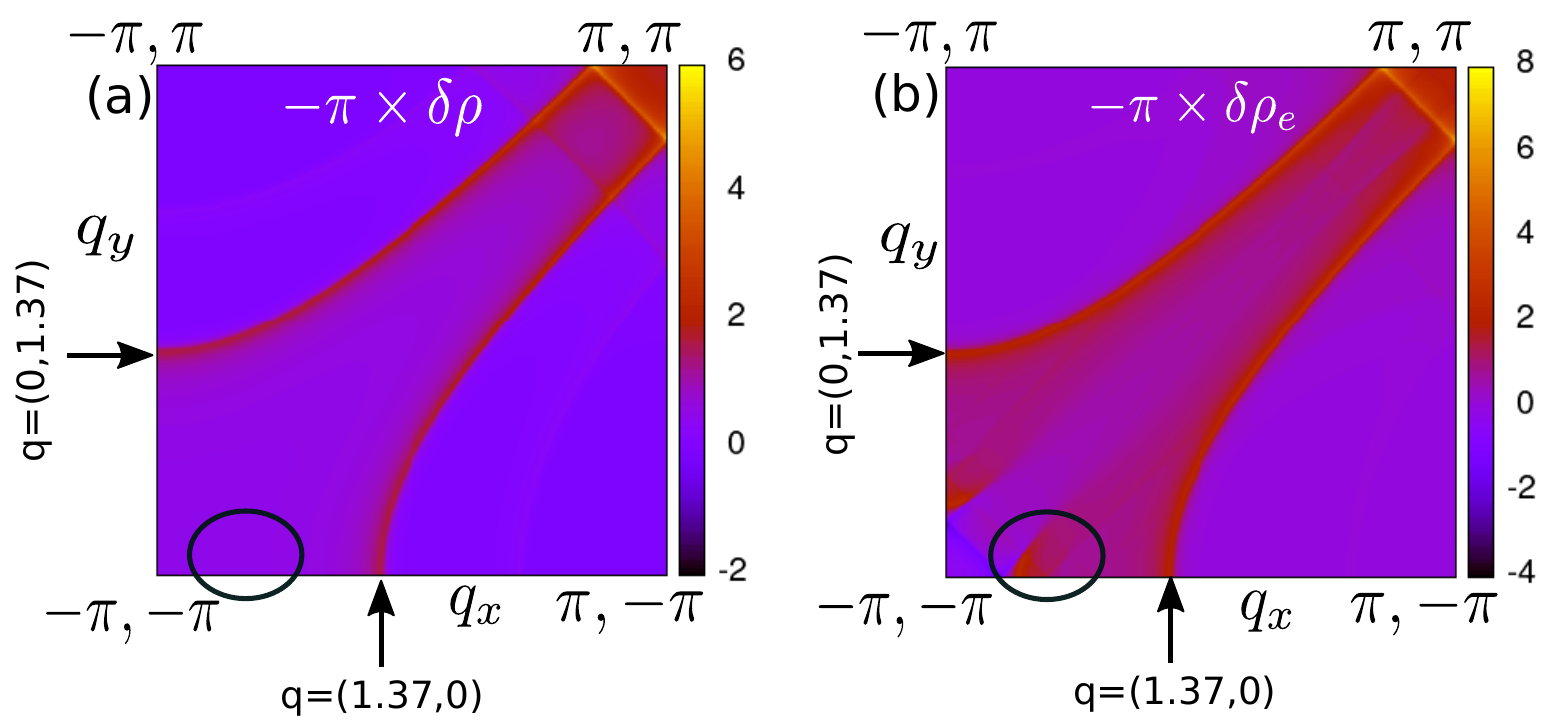} \caption{Color density plot of $\delta\rho(q)$ (a) and $\delta\rho_e(q)$ (b) in the first Brillouin zone for fixed $\omega=0.8$ and $B=0.3$. Black oval indicates a peak which appears only when odd-frequency correlations are ignored. Arrows indicate positions of $q$-values used in Figs.~1 and 2 in the main text.}
\label{fig:momen} 
\end{figure}

In the main text and Fig.~\ref{fig:posnegotherq} of this SM, our analysis is based on fixed $q$-values but varying the frequency/bias $\omega$. Having established the role of odd-frequency correlations in LDOS-change for several fixed $q$, we finally turn our attention to the momentum space structure of the LDOS-change in the full Brillouin zone, now instead fixing $\omega$. In Fig.~\ref{fig:momen} we show the color density maps of the positive bias $\delta\rho(q,\omega)$ and $\delta\rho_e(q,\omega)$ for $\omega=0.8$ and $B=0.3$. In particular, we see that $\delta\rho_e(q,\omega)$ develops additional momentum peaks. For example, one such spurious peak is indicated by black oval in Fig.~\ref{fig:momen}(b). This spurious peak is absent when we appropriately include odd-frequency correlations in $\delta\rho(q,\omega)$ in (a). The momentum structure thus gives another signature of the presence of odd-frequency correlations.

We also note that the peaks around $q=(0,1.37)$ and $q=(1.37,0)$ are present in both $\delta\rho(q,\omega)$ and $\delta\rho_e(q,\omega)$. It is these peaks we show have different $\omega$ dependence in Figs.~1 and 2 of the main text. We have also verified that the $q$-integrated values of $\delta\rho(q,\omega)$ and $\delta\rho_e(q,\omega)$ have similar $\omega$ properties as in Figs.~1 and 2 of the main text, further solidifying our results.

\section{QPI for spin-singlet odd-frequency correlations}

In the main text, we show the LDOS-change for odd-frequency correlations with spin-triplet symmetry. In this section we obtain the LDOS-change for a model which generates spin-singlet odd-frequency correlations in order to fully separate spin and frequency dependence and provide further evidence that the LDOS-change is only tied to the frequency symmetry.

In order to investigate spin-singlet odd-frequency correlations in a transparent and relatable way to the results in the main text, we consider a pair hopping model Hamiltonian on the 2D square lattice originally proposed in the context of cuprates. The Hamiltonian has already been studied in detail in Refs.~\onlinecite{Waardh17,Chakraborty22}. Still, for completeness, we here briefly discuss the model, method, and parameters used. We use this Hamiltonian as an example model which generates spin-singlet odd-frequency correlations without getting into the complexities present if instead using multiband Hamiltonians or heterostructures, which are other systems prone to spin-singlet odd-frequency correlations. The Hamiltonian is given by
\begin{equation}
H_{\rm {PH}}=\sum_{k,\sigma} \xi_{k} c_{k \sigma}^{\dagger} c_{k \sigma} + \sum_{k} \left( \Delta^{Q}_{k} c_{-k+Q/2 \downarrow} c_{k+Q/2 \uparrow} + \textrm{H.c.} \right) + \text{constant},
\label{eq:phHamilmf}
\end{equation}
where the electron dispersion is $\xi_{k}=-2t(\cos(k_x)+\cos(k_y))-\mu$ with $t=1$ being the energy unit and we tune $\mu$ such that the average density of electrons $\rho=\sum_{k,\sigma}\langle c^{\dagger}_{k\sigma}c_{k\sigma} \rangle$  is fixed to $0.8$. $\Delta^{Q}_{k}$ is the spin-singlet superconducting order parameter obtained by the self-consistency relation
\begin{equation}
\Delta^{Q}_k=\sum_{k^{\prime}}V_{k,k^{\prime},Q} \langle c_{k^{\prime}+Q/2 \uparrow}^{\dagger} c_{-k^{\prime}+Q/2 \downarrow}^{\dagger} \rangle.
\label{eq:scph}
\end{equation}
Note that this superconducting order parameter is taken to have a finite center of mass momentum even in the absence of any external magnetic field. This offers a direct connection to the result in the main text, but notably avoid any spin-polarization or spin-triplet formation coupled to applying an external magnetic field.
This finite momentum character comes from the interaction strength, given by 
\begin{equation}
V_{k,k^{\prime},q}=-U\Gamma(q)\left(\gamma(k)\gamma(k')+\eta(k)\eta(k')\right), \label{eq:phint} 
\end{equation}
where $\gamma(k)=\cos(k_x)+\cos(k_y)$ and $\eta(k)=\cos(k_x)-\cos(k_y)$ are the two form factors for nearest-neighbor attraction on a square lattice, and $U$ is a constant attraction strength. The aspect of longer range pair hopping is embedded in the factor $\Gamma(q)$ given by
\begin{equation}
\Gamma(q)= e^{-\frac{\left(q_x-\tilde{Q}\right)^2}{2\kappa_x^2}}+e^{-\frac{\left(q_x+\tilde{Q}\right)^2}{2\kappa_x^2}}, 
\label{eq:gamma}
\end{equation}
where $\kappa_{x}$ denotes the range of the hopping and the modulation $\tilde{Q}=2\pi/P\hat{x}$ with $P=8$. For concreteness, we set $U=2.0$ and $\kappa_x=0.2$.

Due to the $\eta(k)$ and $\gamma(k)$ components in $V_{k,k^{\prime},Q}$, we can decompose $\Delta^{Q}_k$ as $\Delta^{Q}_k=\Delta^{Q}_d\eta(k)+\Delta^{Q}_s\gamma(k)$, with $\Delta^{Q}_d$ being the $d$-wave SC order parameter and $\Delta^{Q}_s$ being the extended $s$-wave SC order parameter \cite{SudboBook}. We then self-consistently solve for $\Delta^{Q}_{d}$ and $\Delta^{Q}_{s}$. In order to obtain the global energy minimum, we here calculate the ground state energy using $E=\sum_{k,\sigma}\xi_{k}\langle c^{\dagger}_{k\sigma}c_{k\sigma} \rangle-(\Delta^{Q}_d)^2/(U\Gamma(Q))-(\Delta^{Q}_s)^2/(U\Gamma(Q))+\mu\rho$ \cite{Waardh17}. We find that the self-consistent value of $\Delta^{Q}_{s}$ is very small for all $Q$ and negligible compared to $\Delta^{Q}_{d}$. The global energy minima occurs for $Q=(0.76,0)$ with the corresponding $\Delta^{Q}_{d}=0.325$. As shown in Ref.~\onlinecite{Chakraborty22}, the Hamiltonian in Eq.~\eqref{eq:phHamilmf} gives odd-frequency spin-singlet $p$-wave correlations due to the formation of a finite momentum superconducting state.

\begin{figure*}[t]
\centering
\includegraphics[width=0.8\linewidth]{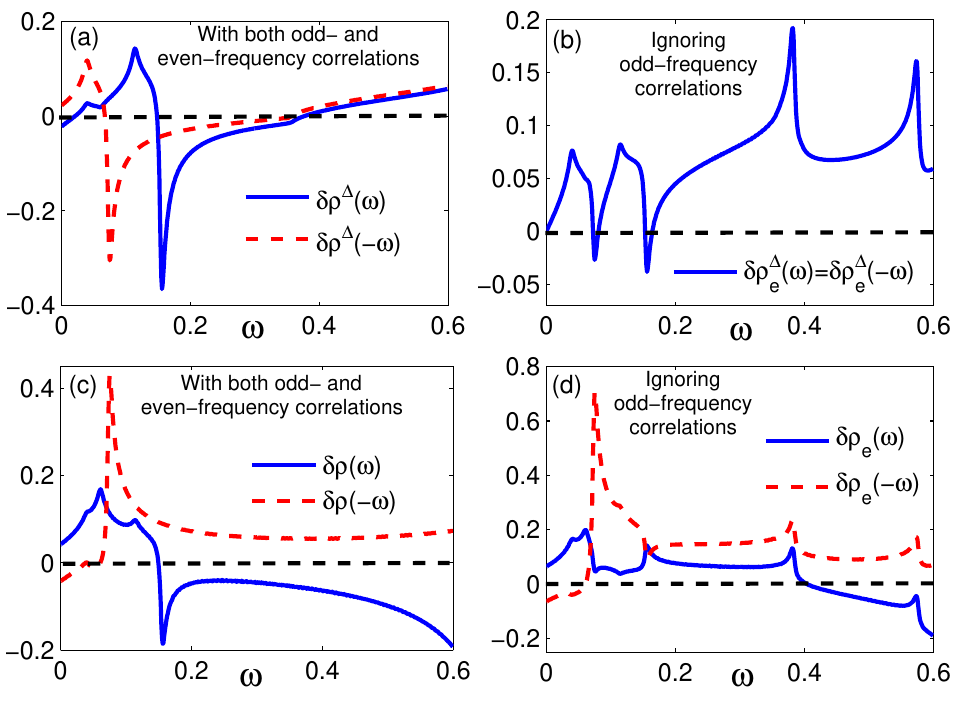} \caption{LDOS-change as a function of $\omega$ for model with spin-singlet odd-frequency correlations in Eq.~\eqref{eq:phHamilmf} for $q=(0.4,0)$. Positive and negative bias values of the superconducting contributions $\delta\rho^{\Delta}$ (a) and $\delta\rho^{\Delta}_{e}$ (b), and total contributions $\delta\rho$ (c) and $\delta\rho_{e}$ (d). Note that no absolute value is taken in order to  display the full sign-structure.}
\label{fig:phLDOSchange} 
\end{figure*}

We next calculate the LDOS-change using the same method as in the main text. In Fig.~\ref{fig:phLDOSchange} we show the LDOS-changes for the Hamiltonian in Eq.~\eqref{eq:phHamilmf}. In (a,b) we first show the contribution for positive and negative bias coming only from the superconducting correlations $\delta\rho^{\Delta}$ and $\delta\rho^{\Delta}_{e}$, respectively. In the presence of both odd- and even-frequency correlations, $\delta\rho^{\Delta}(\pm\omega)$ show clear bias asymmetry in (a). In particular, $\delta\rho^{\Delta}(\omega)$ show positively and negatively valued peaks at very different $\omega$ compared to $\delta\rho^{\Delta}(-\omega)$. This is the first notable feature. In contrast, when ignoring the odd-frequency correlations in $\delta\rho_{e}^{\Delta}(\omega)=\delta\rho_{e}^{\Delta}(-\omega)$, there is no bias asymmetry, as clearly seen in (b). 
The asymmetric position of the peaks in $\delta\rho^{\Delta}(\pm\omega)$ but not in $\delta\rho_{e}^{\Delta}(\pm\omega)$ is in full agreement with the results in the main text in Fig.~2(c,d) but then for a spin-triplet odd-frequency system.
Hence, this asymmetry cannot be described to the spin nature of the system but is entirely due to the odd symmetry in $\omega$ coming from odd-frequency correlations.
We next look at the total LDOS-change $\delta\rho$ and $\delta\rho_{e}$ in Fig.~\ref{fig:phLDOSchange}(c,d). As seen in (c), $\delta\rho(-\omega)$ shows a clear positive peak at $\omega\approx 0.1$ and $\delta\rho(-\omega)$ then remains positive for all $\omega>0.1$. On the other hand, $\delta\rho(\omega)$ shows a positive peak around $\omega\approx 0.1$ but then changes sign to develop a negative peak at $\omega\approx 0.16$ and keeping a negative value for $\omega>0.16$. If we instead insist on ignoring the odd-frequency correlations, $\delta\rho_{e}(\pm\omega)$ in (d) show nearly similar peak positions. Moreover, $\delta\rho_{e}(\pm\omega)$ are both positive with $\delta\rho_{e}(\omega)$ only changing sign at a large $\omega\approx 0.4$. Again, this behavior is very similar to that of Fig.~2(a,b) in the main text.

To summarize, the results in this section show contrasting features of $\delta\rho(\pm\omega)$ only in the presence of odd-frequency correlations. This explicitly demonstrates that this bias asymmetry cannot be tied to the spin degree of freedom as we arrive at similar results for both spin-singlet and spin-triplet odd-frequency correlations.
Taken together, our findings thus give a clear evidence that LDOS-change and QPI can generically be used for experimental determination of odd-frequency correlations, independent of spin symmetries.

\end{document}